\documentclass[aps,pra,twocolumn,showpacs,superscriptaddress]{revtex4-1}
\usepackage{enumerate}
\usepackage{graphicx}
\usepackage{amsmath}
\usepackage{mathtools}
\usepackage{amsfonts}
\usepackage{amssymb}
\usepackage{color}

\newcommand{\ket}[1]{\left| #1 \right\rangle}
\newcommand{\bra}[1]{\left\langle #1 \right|}

\begin{document}
\today
\title{Entangling power of holonomic gates in atom-cavity systems}
\author{Vahid Azimi Mousolou\footnote{Electronic address: v.azimi@sci.ui.ac.ir}}
\affiliation{Department of Mathematics, Faculty of Science, University of Isfahan, Box 81745-163 Isfahan, Iran}
\affiliation{School of Mathematics, Institute for Research in Fundamental Sciences (IPM), 
P. O. Box 19395-5746, Tehran, Iran}
\author{Erik Sj\"oqvist\footnote{Electronic address: erik.sjoqvist@kvac.uu.se}}
\affiliation{Department of Physics and Astronomy, Uppsala University, Box 516, 
SE-751 20 Uppsala, Sweden}
\begin{abstract}
Our goal is to provide a new approach to the construction of geometry-induced entanglement 
between a pair of $\Lambda$ type atoms in a system consists of $N$ identical atoms by means 
of nonadiabatic quantum holonomies. By employing the quantum Zeno effect, we introduce a 
tripod type interaction Hamiltonian between two selected atoms trapped in an optical cavity, 
which allows arbitrary geometric entangling power. This would be a substantial step toward 
resolving the feasibility of realizing universal nonadiabatic holonomic entangling two-qubit gates.   
\end{abstract}
\pacs{}
\maketitle

\section{Introduction}
Environment induced noise and decoherence is a great bane in realizing quantum computers. 
There are diverse theoretical proposals developed for physical implementations, which try to 
avoid noise completely or at least protect against the effect of noise. Decoherence free 
subspaces \cite{zanardi1997a, lidar1998, lidar2003}, dynamical decoupling 
\cite{viola1998, viola1999}, noiseless subsystems \cite{zanardi1997, knill2000}, topological 
\cite{freedman1998, kitaev2003} and geometric \cite{zanardi99,xiang-bin2001,zhu2002,sjoqvist2012} 
approaches, and quantum error-correction methods \cite{shor1995, steane1996} are among 
these proposals.

As one of the key approaches in achieving fault-tolerant quantum computation, 
geometric/holonomic quantum computation has caught a great deal of interest in recent 
years. Holonomic quantum computation was initially introduced in the adiabatic regime 
\cite{zanardi99, ekert2000,duan2001,faoro2003, solinas2003} and subsequently developed 
for nonadiabatic processes 
\cite{xiang-bin2001,zhu2002, zhu2003a, zhu2003b, sjoqvist2012, mousolou2014}, the latter 
being compatible with the short coherence time of quantum bits (qubits). Nonadiabatic 
holonomic gates have been experimentally implemented in various physical settings, such 
as NMR \cite{feng2013, li2017}, superconducting transmon \cite{Abdumalikov2013}, 
NV centers in diamond \cite{Arroyo-Camejo2014, Zu2014, Zhou2017, sekiguchi2017}. 
Further feasible schemes have been established for nonadiabatic geometric processing 
with spin qubit systems \cite{mousolou2017b}, and pseudo-spin charge qubits 
\cite{mousolou2017a}. Moreover, nonadiabatic holonomic quantum computation has 
been incorporated with decoherence free subspaces 
\cite{xu2012, xu2014a, liang2014, xue2015, zhou2015, xue2016, zhao2017, mousolou2018}, 
noiseless subsystems \cite{zhang2014}, and dynamical decoupling \cite{xu2014b}. Nonetheless, 
the construction of an externally controlled multipartite system with full entangling power for 
nonadiabatic holonomic processing is one of the main and challenging obstacles from a practical 
perspective.

In this paper, we discuss an atom-cavity system, which not only allows for combining two 
fault-tolerant methods in quantum computing, namly, decoherence-free subspaces and 
holonomic quantum processing, but also allows for full geometry-induced entangling power. 
The system we present here consists of separated three-level atoms placed at fixed positions 
inside an optical cavity, which can be implemented by using current technology. Taking the 
advantage of the quantum Zeno effect, we establish a tripod interaction in a decoherence-free 
subspace corresponding to two selected atoms in this cavity, each of which represents a qubit 
system. We then demonstrate that this tripod arrangement permits implementation of nonadiabatic 
holonomic two-qubit gates with arbitrary entangling power. The generic nature of the proposed 
scheme would help to overcome the practical challenges in realizing universal 
holonomic quantum information processing.

\section{Atom-cavity system}
\begin{figure}[h]
\begin{center}
\includegraphics[width=7.5 cm]{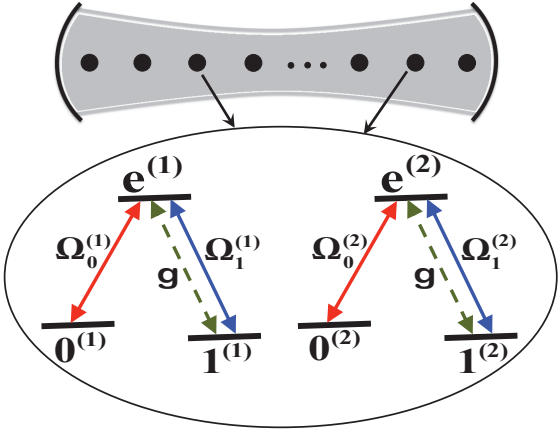}
\end{center}
\caption{(Color online) The upper panel shows a system of identical atoms arranged in a 
line and trapped along the symmetry axis of an optical cavity. The lower panel illustrates 
two selected $\Lambda$ type atoms trapped in the cavity tuned with strength $g$ on  
resonance with the $\ket{1^{(i)}} \leftrightarrow \ket{e^{(i)}}$, $i=1,2$, transitions. The 
desired tripod type interaction shown in Fig.~\ref{fig:tripod} may be achieved by applying 
distinct laser pulses with tuned Rabi frequencies $\Omega_{0}^{(i)}$ and $\Omega_{1}^{(i)}$, 
respectively inducing $\ket{0^{(i)}}\leftrightarrow\ket{e^{(i)}}$ and $\ket{1^{(i)}} \leftrightarrow 
\ket{e^{(i)}}$ transitions in the corresponding atom $i = 1,2$.}
\label{fig:2qubit}
\end{figure}

The system we have in mind consists of $N$ identical atoms, arranged in a line and trapped 
along the symmetry axis of an optical cavity so that each atom can be addressed individually 
(see Fig.~\ref{fig:2qubit}). To create geometry-induced entanglement between atom pairs, 
we further assume that the selected atoms have fixed positions inside the cavity. Without loss 
of generality, we select the atoms fixed at the first and second position in the chain. 

Each atom exhibits a three-level $\Lambda$-type structure, with the atomic ground states 
$\ket{0}$ and $\ket{1}$ coupled to an excited state $\ket{e}$. The ground state levels 
$\ket{0}$ and $\ket{1}$ span a qubit state space. The atomic transitions 
$\ket{0}\leftrightarrow\ket{e}$ and $\ket{1}\leftrightarrow\ket{e}$ are assumed to be in 
resonance with the field mode in the cavity. For simplicity, the atom-cavity coupling constant 
is taken to be $g$ for all atoms. For no photon in the cavity mode, one finds that the 
computational states $\ket{00}, \ket{01}, \ket{10}, \ket{11}$ and the maximally entangled 
trapped state $\ket{\alpha} = ( \ket{1e}-\ket{e1} ) /\sqrt{2}$ of the two atoms span a 
decoherence-free subspace (DFS) with respect to cavity emissions \cite{beige2000a}. 

In order to generate entanglement between the selected atoms, we introduce a mechanism 
to manipulate the states inside the above DFS by means of geometric phases. For this, we 
apply resonant laser pulses addressing each atom individually. The Rabi frequencies of the 
laser pulses inducing $\ket{0^{(i)}} \leftrightarrow \ket{e^{(i)}}$ and $\ket{1^{(i)}} \leftrightarrow 
\ket{e^{(i)}}$ transitions in atom $i=1, 2$, are taken to be $\Omega_{0}^{(i)}$ and 
$\Omega_{1}^{(i)}$, respectively. These frequencies are generally complex-valued. 
Thus, the laser part of the conditional Hamiltonian that describes the dynamics of the 
system is given by 
\begin{eqnarray}
H_{{\rm laser}} = \hbar \sum_{i=1}^{2} \sum_{j=0}^{1} 
\Omega_{j}^{(i)}\ket{j^{(i)}}\bra{e^{(i)}}+{\rm H.c.} 
\end{eqnarray}

\begin{figure}[h]
\begin{center}
\includegraphics[width=5 cm]{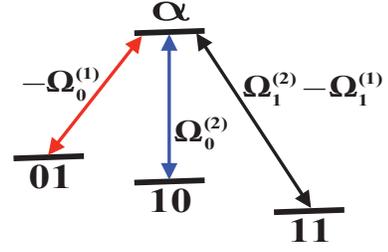}
\end{center}
\caption{(Color online) The conditional tripod interaction picture, where the atoms, the cavity 
mode as well as the environment-induced quantum Zeno effect are taken into account.}
\label{fig:tripod}
\end{figure}

If the amplitude of the Rabi frequencies are much smaller than $\kappa$ and $g^{2}/\kappa$, 
where $\kappa$ is the decay rate of a single photon inside the resonator, then with the help 
of the environment-induced quantum Zeno effect, the system can be kept inside the DFS 
during the evolution of the system \cite{beige2000a, beige2000b}. In this regime, the effective 
Hamiltonian 
\begin{eqnarray}
H_{{\rm eff}} ={\bf P}H_{\text{laser}}{\bf P}, 
\end{eqnarray}
where ${\bf P}$ is the projection operator on the DFS, leads to the following tripod configuration (see Fig.~\ref{fig:tripod})
\begin{eqnarray}
H_{\text{eff}} & = & \frac{\hbar}{\sqrt{2}} \Big( -\Omega_{0}^{(1)} \ket{01} \bra{\alpha} + 
\Omega_{0}^{(2)}\ket{10}\bra{\alpha}  
\nonumber\\
 & & + (\Omega_{1}^{(2)}-\Omega_{1}^{(1)})\ket{11}\bra{\alpha} + {\rm H.c.} \Big) , 
\label{eq:effH}
\end{eqnarray}
where we have used the short-hand notation $\ket{j^{(1)} k^{(2)}} \equiv \ket{jk}$, 
$j,k=0,1$. 

Solving the eigenvalue problem of $H_{{\rm eff}}$, we obtain orthonormal eigenstates
\begin{eqnarray}
\ket{D_1} & = & e^{-i\phi_{2}}\sin\theta\ket{01}+e^{-i\phi_{1}}\cos\theta\ket{10} , 
\nonumber\\
\ket{D_2} & = & e^{-i\phi_{3}}\cos\varphi\ket{x}-\sin\varphi\ket{11} ,
\nonumber\\
\ket{B_{\pm}} & = & \frac{\ket{y} \pm \ket{\alpha}}{\sqrt{2}} , 
\end{eqnarray}
and corresponding eigenenergies $E_{D_1} = E_{D_2} = 0$,  and 
$E_{B_{\pm}} = \pm \frac{\hbar\omega}{\sqrt{2}}$. Here,  
\begin{eqnarray}
\omega & = & \sqrt{|\Omega_{0}^{(1)}|^{2}+|\Omega_{0}^{(2)}|^{2} + 
|\Omega_{1}^{(2)}-\Omega_{1}^{(1)}|^{2}} , 
\nonumber \\
\Omega_{0}^{(1)} & = & \omega e^{i\phi_1} \sin\varphi \cos\theta , 
\nonumber \\
\Omega_{0}^{(2)} & = & \omega e^{i\phi_2} \sin\varphi \sin\theta , 
\nonumber \\ 
\Omega_{1}^{(2)}-\Omega_{1}^{(1)} & = & \omega e^{i\phi_3} \cos\varphi ,  
\nonumber \\
\ket{x} & = & e^{i\phi_{2}}\sin\theta\ket{10}-e^{i\phi_{1}}\cos\theta\ket{01}, 
\nonumber \\
\ket{y} & = & \sin\varphi \ket{x} + e^{i\phi_3} \cos\varphi \ket{11}.
\end{eqnarray}
These parameters are kept constant for the duration $[0,\tau]$ of the laser pulses, resulting 
in the time evolution operator of the DFS 
\begin{eqnarray} 
\mathcal{U}(\tau ,0) & = & e^{-\frac{i}{\hbar}\int_{0}^{\tau} H_{{\rm eff}} dt} = 
\ket{D_{1}}\bra{D_{1}} + \ket{D_{2}}\bra{D_{2}} 
\nonumber\\
 & & + \cos a_{\tau} \big( \ket{y}\bra{y} + \ket{\alpha}\bra{\alpha} \big)  
\nonumber \\ 
 & & - i\sin a_{\tau} \big( \ket{y}\bra{\alpha} + \ket{\alpha}\bra{y} \big) ,  
\label{eq:TE}
\end{eqnarray}
where $a_{\tau} = \frac{\omega\tau}{\sqrt{2}}$.

\section{Quantum holonomies and geometry-induced entanglement}
By choosing the run time $\tau = \sqrt{2} \pi/\omega$ such that $a_{\tau} = \pi$, the three 
dimensional subspace ${\rm Span} \{ \ket{01},\ket{10},\ket{11} \}$ undergoes cyclic evolution 
in the four dimensional part ${\rm Span} \{ \ket{01},\ket{10},\ket{11},\ket{\alpha} \}$ of the 
DFS, while the remaining two-qubit state $\ket{00}$ is fully decoupled. Moreover, 
along this evolution we would have  
\begin{eqnarray}
\mathcal{U}(t,0) {\bf P}_c  \, \mathcal{U}^{\dagger} (t,0) H_{\text{eff}} \,
\mathcal{U}(t,0) {\bf P}_c \, \mathcal{U}^{\dagger} (t,0) = 0
\end{eqnarray}
for the projection ${\bf P}_c = \ket{01}\bra{01}+\ket{10}\bra{10}+\ket{11}\bra{11}$. In other 
words, the three dimensional subspace ${\rm Span} \{\ket{01}, \ket{10}, \ket{11}\}$ evolves 
along a loop $\mathcal{C}$ in the Grassmannian $G(4,3)$, i.e., the space of three dimensional 
subspaces of the four dimensional DFS, along which the dynamical phase vanishes 
\cite{anandan88}. It follows that  
\begin{eqnarray}
U(\mathcal{C})={\bf P}_c \, \mathcal{U}(\tau ,0) {\bf P}_c
\label{NQH}
\end{eqnarray}
is the nonadiabatic quantum holonomy of the loop $\mathcal{C}$ in the Grassmannian 
$G(4,3)$ \cite{anandan88}. 

It is instructive to compare the present scheme with the tripod-based single-qubit architecture 
for adiabatic geometric manipulation proposed in Ref.~\cite{duan2001}. In addition to being 
based on adiabatic evolution, the loop corresponding to each geometric single-qubit gate 
in Ref.~\cite{duan2001} resides in the Grassmannian $G(3,2)$ since the qubit levels are 
encoded in the two dark states evolving in the three dimensional space ${\rm Span} 
\{ \ket{01},\ket{10},\ket{11} \}$.

Since the computational basis state $\ket{00}$ does not contribute to the dynamics described 
by the effective Hamiltonian in Eq.~(\ref{eq:effH}), it remains unchanged during the evolution. 
Therefore, the evolution results in the following two-qubit nonadiabatic holonomic gate 
\begin{eqnarray}
U = \ket{00}\bra{00}+U(\mathcal{C}),
\end{eqnarray}
which takes the following form in the computational ordered basis 
$\{\ket{00}, \ket{01}, \ket{10}, \ket{11}\}$
\begin{widetext}
\begin{eqnarray}
U=\left(
\begin{array}{cccc}
 1 & 0  & 0&0  \\
 0 & 1-2\sin^{2}\varphi\cos^{2}\theta & e^{-i\phi_{21}}\sin 2\theta\sin^{2}\varphi & 
 e^{-i\phi_{31}}\sin 2\varphi\cos\theta  \\ 
 0 &  e^{i\phi_{21}}\sin 2\theta\sin^{2}\varphi  & 1-2\sin^{2}\varphi\sin^{2}\theta & 
 -e^{-i\phi_{32}}\sin 2\varphi\sin\theta \\
 0 &  e^{i\phi_{31}}\sin 2\varphi\cos\theta & -e^{i\phi_{32}}\sin 2\varphi\sin\theta & 
 -\cos 2\varphi
\end{array}
\right),
\end{eqnarray}
\end{widetext}
where $\phi_{lk}=\phi_{l}-\phi_{k},\ l,k=1, 2, 3$.

An important feature of the above two-qubit gate $U$ is that it provides geometric gates 
with arbitrarily large entangling power. To see this, let us look at some entangling 
characteristics of $U$. Evaluating the local invariances \cite{makhlin2002}, one obtains  
\begin{eqnarray}
G_{1} & = & -\sin^8 \varphi \sin^4 2\theta , 
\nonumber \\
G_{2} & = & \cos 2\varphi + 2\sin^{2}\varphi \left( \cos^{2}\varphi + 
\cos 4\theta \sin^{2}\varphi \right) ,
\label{eq:localinvariants}
\end{eqnarray}
which consequently results in the entangling power \cite{zanardi00-r,  balakrishnan10}
\begin{eqnarray}
e_{p}(U) & = & \frac{2}{9} \left( 1-\left| G_1 \right| \right) 
\nonumber \\
 & = & \frac{2}{9} \left( 1-\sin^{8}\varphi\sin^{4}2\theta \right) .
\label{eq:ep} 
\end{eqnarray}

\begin{figure}[h]
\begin{center}
\includegraphics[width=7 cm]{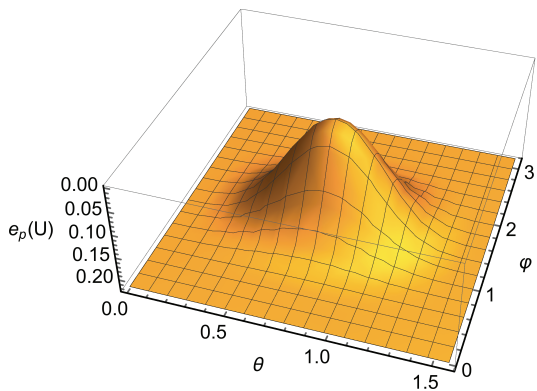}
\end{center}
\caption{(Color online) Entangling power, $e_{p}(U)$, of the two-qubit nonadiabatic holonomic 
gate $U$ as a function of the control parameters $\theta$ and $\varphi$ in a single period.}
\label{fig:ep}
\end{figure}

As shown in Fig.~\ref{fig:ep}, a careful tuning of the laser pulses can provide any entangling 
power.

From Eqs.~(\ref{eq:localinvariants}) and (\ref{eq:ep}), one may note that the entangling 
nature of the gate $U$ does not in general depend on the complex nature of the Rabi 
frequencies $\Omega_{0}^{(i)}$ and $\Omega_{1}^{(i)}$, $i=1,2$. Extracting the corresponding 
symmetry reduced geometric coordinate $\left( c_1, c_2, c_3 \right)$ of $U$ on the Weyl 
chamber \cite{zhang2003}, which classifies non-local two-qubit gates, we have 
\begin{eqnarray}
\left( c_1, c_2, c_3 \right) = \left( \frac{\pi}{2}, c, c \right) ,
\end{eqnarray}
where
\begin{eqnarray}
c=\arcsin \left( 2\left| \Omega_{0}^{(1)} \Omega_{0}^{(2)} \right| \omega^{-2} \right).
\end{eqnarray}
This indicates that the geometric gate $U$ covers the whole equivalence classes of two-qubit 
gates along the line segment connecting the equivalence class of special perfect entanglers 
[CNOT], represented by the coordinate $\left( \frac{\pi}{2}, 0, 0 \right)$, to the class of local 
gates represented by the coordinate $\left( \frac{\pi}{2}, \frac{\pi}{2}, \frac{\pi}{2} \right)$ on 
the Weyl chamber. Moreover, if the lasers are tuned so that 
\begin{eqnarray}
\left| \Omega_0^{(1)} \Omega_0^{(2)} \right| = \frac{\omega^2}{2\sqrt{2}} , 
\end{eqnarray}
then the geometric gate $U$ belongs to the equivalence class of perfect entanglers 
corresponding to the point $\left( \frac{\pi}{2}, \frac{\pi}{4}, \frac{\pi}{4} \right)$ on the Weyl 
chamber. The entangling nature in fact depends on the absolute frequency ratios 
$\left| \Omega_{0}^{(2)} / \Omega_{0}^{(1)} \right|$ and 
$\left| \big( \Omega_1^{(2)} - \Omega_1^{(1)} \big) / \omega \right|$. The gate $U$ tends 
to the equivalence class of special perfect entanglers, denoted as [CNOT], with maximum 
entangling power of $\frac{2}{9}$, when $\left| \Omega_0^{(2)} / \Omega_0^{(1)} \right| \rightarrow 
0, \infty$ or $\left| \big( \Omega_1^{(2)} - \Omega_1^{(1)} \big) /\omega \right| \rightarrow 1$. 
Table \ref{U-classes} specifies some frequencies to achieve different class of entangling gates.

\begin{widetext}
\begin{center}
\begin{table}[htp]
\caption{Entanglement characteristics of the geometric two-qubit entangling gate $U$ for 
some specific frequencies}
\vskip 0.3 cm 
\begin{center}
\begin{tabular}{|c|c|c|c|c|c|}
\hline
 & & & & & \\ 
$\#$ & $(\Omega_{0}^{(1)},\ \Omega_{0}^{(2)},\ \Omega_{1}^{(2)}-\Omega_{1}^{(1)})$ 
 & $G_{1}$ & $G_{2}$ & $e_{p}(U)$ & Weyl chamber \\
 & & & & & coordinate\\
\hline 
1 & $(0, 0, \ne 0)$ & 0 & 1 & 2/9 & $\left( \pi /2, 0, 0 \right) = $ [CNOT] \\
\hline
2 & $(0, \ne 0, 0)$ & 0 & 1 & 2/9 & $( \pi/2, 0, 0 ) = $ [CNOT] \\
\hline
3 & $(\ne 0, 0, 0)$ & 0 & 1 & 2/9 & $( \pi/2, 0, 0 ) = $ [CNOT] \\
\hline
4 & $(0, \ne 0, \ne 0)$ & 0 & 1 & 2/9 & $( \pi/2, 0, 0 ) = $ [CNOT] \\
\hline
5 & $(\ne 0, 0, \ne 0)$ & 0 & 1 & 2/9 & $( \pi/2, 0, 0 ) = $ [CNOT] \\
\hline
6 & $(\ne 0, \ne 0, 0)$ & $ -\sin^{4}2\theta$ & $2\cos 4\theta-1$ & 
$(2/9) (1-\sin^{4} 2\theta)$ & $(\pi/2, 2\theta, 2\theta)$,\ \ $0<\theta\le\frac{\pi}{4}$ \\
\hline
7 & $(\ne 0, \ne 0, \ne 0$) & $-\sin^{8}\varphi$ & $1-4\sin^{4}\varphi$ & 
$\frac{2}{9}(1-\sin^{8}\varphi)$ & $(\pi/2, c, c)$ \\
 & & & & & $0 < c = \arcsin(\sin^{2}\varphi) \le \pi/2$ \\
\hline
\end{tabular}
\end{center}
\label{U-classes}
\end{table}
\end{center}
\end{widetext}

Note that the tripod configuration in Fig.~\ref{fig:tripod} reduces to a two level interaction 
system for the three first upper cases in the table \ref{U-classes}. Therefore, in these cases, 
the loop $\mathcal{C}$ would effectively reside in the Grassmannian G(2, 1) and its 
corresponding nonadiabatic quantum holonomy given in Eq.~(\ref{NQH}) would describe 
only an Abelian nonadiabatic geometric phase \cite{aharonov1987}. However, in the other 
cases listed in the table, the tripod structure reduces to a three-level $\Lambda$ structure, 
which would instead correspond to the effective loop $\mathcal{C}$ reside in the Grassmannian 
G(3, 2) an the accompanying non-Abelian quantum holonomy. In other words, 
Tab.~\ref{U-classes} shows that perfect geometry-induced entanglement can be achieved 
through both Abelian and non-Abelian quantum holonomies in the above proposed 
interaction picture. 

One may notice that the approach in Ref. \cite{beige2000a} is a special example of the 
case listed in row four of the table \ref{U-classes}. The present work, in other words, is an 
expansion of the proposal in Ref.~\cite{beige2000a} introducing a wider class of entangling 
gates with more freedom in the choice of frequencies. Our analysis shows that nonadiabatic holonomies have full entangling power.

\section{Conclusions}
We have followed the nonadiabatic geometric approach to study the entangling power of 
quantum holonomy in an atom-cavity system. We have established a nonadiabatic holonomic 
manipulation of two decoherence-free qubits, described in terms of quantum 
Zeno effect in the study of a chain of identical atoms trapped in an optical cavity. We achieved 
arbitrary geometry-induced entangling power through the proposed nonadiabatic holonomic 
approach. Moreover, the proposed system benefits from both decoherence-free subspace 
and holonomic manipulation methods to gain robustness to decoherence effects and parameter noises, respectively, and to introduce an efficient 
way of entangling qubit systems. Our scheme is generic, scalable, and can be implemented 
in a wide range of atomic and ionic systems trapped in cavities \cite{pachos2001,lukin2003}.

\section{Acknowledgment }
This work was supported by Department of Mathematics at University of Isfahan (Iran). V.A.M. acknowledges financial support from the Iran National Science Foundation (INSF) through Grant No. 96008297. E.S. acknowledges financial support from the Swedish Research Council (VR) through Grant No. 2017-03832.

\bibliography{NHQCQD}
\end{document}